\documentclass[journal,twoside,final]{IEEEtran}
\usepackage{amsfonts}
\usepackage{amssymb}
\usepackage{amsmath}
\usepackage[dvips]{graphicx}
\usepackage{cite}

\usepackage{enumitem}
\usepackage{balance}
\usepackage{multirow}
\usepackage{balance}
\usepackage{fixmath}
\usepackage{balance}
 \usepackage{algpseudocode}
  \usepackage{algorithm}
\newtheorem{remark}{Remark}

\begin{document}

\title{Wireless Powered Communications with Non-Orthogonal Multiple Access}
\author{
Panagiotis D. Diamantoulakis,~\IEEEmembership{Student Member,~IEEE,}
Koralia N. Pappi,~\IEEEmembership{Member,~IEEE,}\newline
Zhiguo Ding,~\IEEEmembership{Member,~IEEE,} and George K. Karagiannidis,~\IEEEmembership{Fellow,~IEEE,}
\thanks{P. D. Diamantoulakis, K. N. Pappi, and G. K. Karagiannidis are with the Department of Electrical and Computer Engineering, Aristotle University
of Thessaloniki, 54 124, Thessaloniki, Greece (e-mails: \{padiaman,kpappi,geokarag\}@auth.gr)}
\thanks{Zhiguo Ding is with the School of Computing and Communications, Lancaster University, LA1 4WA, UK (e-mail: z.ding@lancaster.ac.uk).}
\thanks{The work of P. D. Diamantoulakis and G. K. Karagiannidis was funded by the NPRP grant \# NPRP 6-1326-2-532 from the Qatar National Research Fund (a member of Qatar Foundation). The statements made herein are solely the responsibility of the authors.}
\thanks{Part of this work has been accepted for publication at the IEEE International Conference on Communications (ICC), Kuala Lumpur, Malaysia, May 2016.}

}


\maketitle

%
%

\begin{abstract}
We study a wireless-powered uplink communication system with non-orthogonal multiple access (NOMA), consisting of one base station and multiple energy harvesting users. More specifically, we focus on the individual data rate optimization and fairness improvement and we show that the formulated problems can be optimally and efficiently solved by either linear programming or convex optimization. In the provided analysis, two types of decoding order strategies are considered, namely \textit{fixed decoding order} and \textit{time-sharing}. Furthermore, we propose an efficient greedy algorithm, which is suitable for the practical implementation of the time-sharing strategy. Simulation results illustrate that the proposed scheme outperforms the baseline orthogonal multiple access scheme. More specifically, it is shown that NOMA offers a considerable improvement in throughput, fairness, and energy efficiency. Also, the dependence among system throughput, minimum individual data rate, and harvested energy is revealed, as well as an interesting trade-off between rates and energy efficiency. Finally, the convergence speed of the proposed greedy algorithm is evaluated, and it is shown that the required number of iterations is linear with respect to the number of users.
\end{abstract}


\section{Introduction}\label{S:Intro}
\IEEEPARstart{A}{}major limitation of untethered communication equipments is that devices operate for a finite duration, which is limited by the lifetime of batteries \cite{Sude}. To this end, energy harvesting (EH), which refers to harnessing energy from the environment or other energy sources and converting it to electrical energy, has recently received a lot of attention. Apart from offering a promising solution for energy-sustainability of wireless nodes in communication networks \cite{harvesting}, EH also reduces considerably the operational expenses \cite{Sude}. 

An alternative to traditional energy harvesting, relying on natural energy sources (e.g. solar power), is wireless power transfer \cite{Grover, Varshney}. 
Particularly, wireless signals can be used for simultaneous wireless information and power transfer (SWIPT). In this framework, nodes use the power by the received signal to charge their batteries \cite{Kwan}, or to transmit the information to a base station (BS) \cite{Ju}. However, in practice, nodes cannot harvest energy and receive/transmit information simultaneously \cite{Trung1, Suraweera, Ju, krik3}. In order to overcome this difficulty, two strategies have been proposed, i.e power splitting and time switching \cite{Xun, krik3}. The idea of SWIPT has been studied in various case studies, such as one source-destination pair \cite{Kwan}, multiple-input multiple-output (MIMO) communications systems \cite{Xiang, krik2}, orthogonal frequency division multiple access (OFDMA) \cite{OFDMA}, and cooperative networks \cite{Esnaola, krik1, Panos}.

Among the proposed SWIPT applications, this paper focuses on the joint design of downlink energy transfer and uplink information transfer in multiuser communication systems, which has been initially studied in \cite{Ju}. Taking into account the time-sharing technique, the authors in \cite{Ju} have proposed a novel protocol referred to as \textit{harvest-then-transmit}, where the users first harvest energy, and then they transmit their independent messages to the BS by using the harvested energy. More specifically, it was assumed that the users utilize time division multiple access (TDMA) for information transmission. 

\subsection{Motivation}
Although relying on the harvested energy for transmission has many benefits, it has a negative impact on the individual data rates achieved by the EH nodes. Consequently, existing methods, which increase power-bandwidth efficiency, should be carefully explored \cite{Matthaiou, Xiaoming}. Toward this direction, the utilization of orthogonal multiple access schemes, such as TDMA, might not be the most appropriate choice.

On the other hand, non-orthogonal multiple access (NOMA) was proved to increase spectral efficiency \cite{Saito}. For this reason, it has been recently proposed for LTE Advanced \cite{3gpp}, in which it is termed as multi-user superposition transmission (MUST). Furthermore, it has also been recognized as a promising multiple access technique for fifth generation (5G) networks \cite{DOCOMO, Ding3, VLC, Tafazoli}. NOMA is substantially different from orthogonal multiple access schemes, i.e. time/frequency/code division multiple access schemes, since its basic principle is that the users can achieve multiple access by exploiting the power domain. For this reason, the decoder needs to implement a joint processing technique, such as successive interference cancellation (SIC). 

The performance of a downlink NOMA scheme with randomly deployed users has been investigated in \cite{Ding3}, while the application of NOMA for the downlink of cooperative communication networks was proposed in \cite{Ding2}, among others. Also, in \cite{Tafazoli}, the authors study NOMA for the uplink of a communication network, consisting of traditional nodes with fixed energy supplies. However, when NOMA is combined with wireless powered communications, the capacity region is strongly affected by the amount of the harvested energy. This is because NOMA uses the power field to achieve multiple access. Consequently, rate maximization and user fairness are still open problems.

\subsection{Contribution}
Unlike recent literature, in this work, we study the application of NOMA for a wireless-powered uplink communication system, which consists of one BS and multiple energy harvesting users, in order to increase the individual data rates and the user fairness. Note that the implementation of NOMA in the uplink is not a burden for the users, since the encoding complexity at the users' side is not affected, while their synchronization is usually simpler than the case of TDMA. For this purpose, we optimize the related variables, taking into account two different criteria: the sum-throughput and the equal individual data rate maximization. The corresponding contribution is summarized as follows:
\begin{itemize}
\item While the sum-throughput is maximized, further improvement of the minimum individual data rate among users is achieved. 
\item We optimize the time used for energy harvesting and the time-sharing variables related to SIC. 
\item Regarding equal individual data rate maximization when the time-sharing technique is utilized, we provide a tractable reformulation of the initial optimization problem.
\item We show that all formulated problems can be optimally solved by either linear programming  or convex optimization tools, which is important for the practical implementation of the proposed scheme. 
\item We propose a greedy algorithm for the optimization of the variables related to the time-sharing technique, which is very efficient, in terms of performance and convergence speed.
\end{itemize}

Extended simulation results illustrate that the application of the proposed NOMA scheme has the following advantages, when compared to the case of TDMA: i) it leads to a notable increase of the minimum individual data rate, and/or, ii) it improves fairness. Finally, an interesting trade-off between the time used for energy harvesting and information transmission is revealed, as well as the dependence among system throughput, minimum individual data rate, energy efficiency, and harvested energy.

\subsection{Structure}
The rest of the paper is organized as follows. Section II
describes the considered communication and energy harvesting
model, while it defines the user individual data rate and the system sum-throughput. The optimization problem of system sum-throughput maximization is formulated and solved in section III. Also, in the same section, the impact of the decoding order of the users' messages on the individual data rates is discussed, both theoretically and with specific illustrative examples, while a greedy algorithm regarding the time-sharing configuration is proposed. The optimization problems of equal individual data rate maximization using fixed decoding order and time-sharing configuration are formulated and solved in sections IV and V, respectively. Section VI presents and discusses the simulation results and finally, section VII concludes the paper with some remarks.

\section{System Model}
\begin{figure}
\centering
\includegraphics[width=\linewidth]{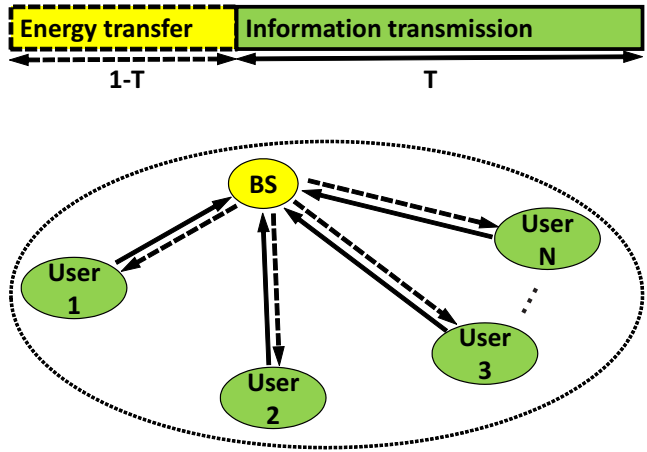}
\caption{Sequential energy transfer and information transmission in NOMA communication networks.}
\label{fig1}
\end{figure}

We consider a wireless network consisting of $N$ users and one BS, where all are equipped with a single antenna. The path loss factor from the BS to user $n$ is denoted by $\mathcal{L}_{0n}$, while the channel coefficient is given by the complex random variable $h_{0n}\sim\mathcal{CN}(0,1)$. The communication is divided into time frames of unitary duration, and it is assumed that the channel state remains constant during a time frame, and can be perfectly estimated by the BS. The considered system model is presented in Fig.\ref{fig1}.

\subsection{Harvest-then-Transmit Protocol}
We consider that  the network adopts a harvest-then-transmit protocol, i.e. at first, the amount of time $1-T, 0\leq T \leq 1$ is assigned to the BS to broadcast wireless energy to all users \cite{Ju}. The remaining time, $T$, is assigned to users, which simultaneously transmit their independent information to the BS by using the energy harvested from the first phase. In order to detect the users' signals, the BS implements a joint processing technique \cite{Ding3, honig}, and for this purpose, it employs NOMA \cite{Tafazoli}. We assume that the energy transmitted by each user $n$ is limited by the amount of harvested energy, i.e. during time portion $T$, each user can only use the energy that was harvested during $1-T$. The energy harvested by the $n$-th user is
\begin{equation}
E_n=\mathcal{G}_{0}\mathcal{G}_{n}\eta_1{L}_{0n}|h_{0n}|^2P_0(1-T),
\end{equation}
where $\mathcal{G}_{0}$ and $\mathcal{G}_{n}$ are the directional antenna gains of the BS and the $n$-th user, respectively,  $0<\eta_1<1$ is the energy harvesting efficiency,  and $P_0$ is the transmit power of the BS. The transmit power of the $n$-th user is given by
\begin{equation}
P_n=\frac{E_n}{T}.
\end{equation}
\subsection{Optimization Objectives}
In this paper, two distinct objectives are set, for optimizing the provided quality-of-service (QoS), i.e. system performance and user fairness. These objectives are described below.

\emph{Maximization of the achievable system throughput (non-symmetric rates):} When this objective is set, users are allowed to transmit with non-symmetric individual rates and thus we seek to maximize the sum-capacity of the users, i.e. optimize the achievable rate region of the network, so that it contains the points which correspond to the maximum system throughput. In order to increase user fairness, we also seek to maximize the individual data rate of the weakest user, given that the achievable system throughput is first maximized. The system throughput will be denoted by $\mathcal{R}_{\mathrm{tot}}$.

\emph{Maximization of the equal individual data rates (symmetric rates):} When the users transmit with equal individual data rates, their data rate corresponds to the minimum achievable throughput among the users. Thus, when this objective is set, we seek to optimize the achievable rate region of the network, so that it contains those points that maximize the achievable throughput of the weakest user, without necessarily seeking to maximize the achievable system throughput. The equal individual data rate will be denoted by $\mathcal{R}_{\mathrm{eq}}$.

Note that the above objectives are not equivalent, since the achievable system throughput might be maximized at the expense of the minimum individual data rate and vice versa.

In the above cases, the sum-rate of the network, denoted by $\mathcal{R}_{\mathrm{sum}}$, is the sum of the individual data rates of the users. When maximizing the system throughput, users are allowed to transmit with asymmetric rates, thus achieving the system throughput. In this case, $\mathcal{R}_{\mathrm{sum}}=\mathcal{R}_{\mathrm{tot}}$. When maximizing the equal individual data rate, all users transmit with symmetric rates, that is, with rate $\mathcal{R}_{\mathrm{eq}}$. In that case, $\mathcal{R}_{\mathrm{sum}}=N\mathcal{R}_{\mathrm{eq}}$.

Alongside the optimization of the aforementioned QoS criteria, we take into account two different approaches concerning the decoding order of the users' messages: i) fixed decoding order and ii) time-sharing, which will both be described in the subsections that follow. The resulting optimization problems can be classified into the following four schemes, which will be referred to as (a)-(d) hereafter:
\begin{enumerate}[label=(\alph*)]
\item Achievable system throughput maximization and minimum individual data rate optimization with fixed decoding order.
\item Achievable system throughput maximization and minimum individual data rate optimization with time-sharing.
\item Equal individual data rate maximization with fixed decoding order.
\item Equal individual data rate maximization with time-sharing.
\end{enumerate}

It should be noted that the time-sharing technique that is used in the schemes (b) and (d) improves the QoS at the expense of higher computational complexity. Since time-sharing assumes many different decoding orders of the users, its maximum complexity depends on the number of all different permutations of the users, as it will be described below. Thus, a balance between optimality and efficiency must be achieved when selecting the number of distinct decoding orders that will be used for time-sharing.

\subsection{Achievable User Throughput in the Case of Fixed Decoding Order}
Next, the achievable user throughput is defined assuming that the users' messages are decoded in an increasing order of their indices. It is worth pointing out that different decoding order does affect achievable user throughput, and this will be discussed in the next subsection. Therefore, for decoding the first user's message ($n=1$), interference is created due to all other users $n=2,...,N$, while on the second user's message, interference is created due to users $n=3,...,N$, and so on. Then, the achievable throughput of the $n$-th user, $1\leq n \leq (N-1)$, denoted by $R_n$ in the case of fixed decoding order, is given by \cite{Tafazoli}
\begin{equation}
\begin{split}
R_n=&T\log_2\left(1+\frac{P_ng_n}{\sum_{j=n+1}^N (P_jg_j)+N_0}\right)\\
=&T\log_2\left(1+\frac{\frac{\eta\rho(1-T) g_n}{T}}{\frac{\eta\rho(1-T)\sum_{j=n+1}^Ng_j}{T}+1}\right),
\end{split}
\label{user rate}
\end{equation}
while the achievable throughput of the $N$-th user is 
\begin{equation}
R_N=T\log_2\left(1+\frac{\eta\rho(1-T)g_N}{T}\right).
\label{N user rate}
\end{equation}
In (\ref{user rate}) and (\ref{N user rate}), $\rho=\frac{P_0}{N_0}$, $\eta=\eta_1\eta_2$, with $\eta_2$ being the efficiency of the user's amplifier, and $N_0$ is the power of the additive white gaussian noise (AWGN). Also, assuming channel reciprocity, $g_n$ is given by $g_n=\mathcal{G}_{0}^2\mathcal{G}_{n}^2{L}_{0n}^2|h_{0n}|^4$.

\subsection{Achievable User Throughput in the Case of Time-Sharing}
The basic principle of time-sharing is that the order of decoding for the users can change for specific fractions of the duration $T$.  In contrast to the case of fixed decoding order, where the users' messages are decoded in an increasing order of their indices, the order of decoding depends on time-sharing \cite{honig}. Next, we propose a simple configuration to realize the time-sharing technique. In general, there are $N!$ configurations with different decoding order, which we call permutations. Let $\tau_m$, with $\sum_{m}\tau_m=1$, denote the portion of time $T$ for which the BS decodes the users' messages, according to the $m$-th permutation. Hereinafter, $\mathbold{\tau}$ denotes the set of values of $\tau_m\forall m$. 

For mathematical clarity, let $\mathbf{A}$ be the matrix, which represents the set of specific $M\leq N!$ permutations, with elements $\mathbf{A}(m,j_{m,n})$, corresponding to the indices of the users, i.e. $\mathbf{A}(m,j_{m,n})=n$. The decoding order of the users during the $m$-th permutation is determined by the indices of the columns, $j_{m,n},\forall n$, for the $m$-th row of matrix $\mathbf{A}$, i.e. if $j_{m,n}<j_{m,l}$, the message of the $n$-th user will be decoded before the message of the $m$-th. More specifically, the value of a matrix  element is the index of a user. The index of the row denotes a specific  permutation, and the index of the column denotes the decoding order of the user in that permutation. For example, if $\mathbf{A}(2,4)=3$, it means that, when the 2-nd permutation is applied, the message of the $3$-rd user will be decoded in the $4$-th order. 

Thus, taking the time-sharing configuration into account, the achievable throughput of the $n$-th user, denoted by $\tilde{R}_n$ in the case of time-sharing, can be written as
\begin{equation}
\begin{split}
\tilde{R}_n&(T)=\\
&\sum_{m=1}^{M}\tau_lT\log_2\left(1+\frac{\frac{\eta\rho(1-T) g_n}{T}}{\frac{\eta\rho(1-T)\sum_{j_{m,k}>j_{m,n}}g_{A(m,j_{m,k})}}{T}+1}\right).
\label{with sharing}
\end{split}
\end{equation}

\subsection{Achievable System Throughput}
Interestingly, the decoding order does not affect the achievable system throughput in NOMA uplink, and any arbitrary decoding order can be assumed to define the system sum-throughput. Thus, taking into account (\ref{user rate}) and (\ref{N user rate}) the achievable system throughput achieved by NOMA is given by \cite{Tafazoli}
\begin{equation}
\begin{split}
\mathcal{R}_{\mathrm{tot}}=&\sum_{n=1}^NR_n=T\sum_{n=1}^{N-1}\left(\log_2\left(\eta \rho\sum_{i=n}^{N}g_i+\frac{T}{1-T}\right)\right.\\
&-\left.\log_2\left(\eta \rho\sum_{i=n+1}^{N}g_i+\frac{T}{1-T}\right)\right)\\
&+T\left(\log_2\left(\eta \rho g_N+\frac{T}{1-T}\right)-\log_2\left(\frac{T}{1-T}\right)\right)\\
&=T\log_2\left(1+\frac{\eta \rho\sum_{n=1}^{N}g_n}{\frac{T}{1-T}}\right).
\end{split}
\end{equation}

\section{System Throughput Maximization and Minimum Throughput Improvement}
In this section, first, the achievable system throughput maximization problem is formulated and solved. Then, elaborating on this solution we improve the minimum individual data rate, considering the schemes (a) and (b).  Thereafter, in order to reduce the complexity of scheme (b), we propose a greedy algorithm, which efficiently optimizes the time-sharing configuration. Finally, we provide an illustrative example, which gives further insight on the nature of the two schemes, taking into account the effect of the distances between the users and the BS. 

\subsection{Achievable System Throughput Maximization}
It can be easily observed that, when $T=0$ or $T=1$, no time or no energy, respectively, is available to the users in order to transmit, and thus the system throughput is zero. The optimization problem, which aims at maximizing the system throughput, can be written as
\begin{equation}
\begin{array}{ll}
\underset{T}{\text{\textbf{max}}}& \mathcal{R}_{\mathrm{tot}} \\
&\mathrm{C}:0<T<1.
\end{array}
\label{system throughput}
\end{equation}
In (\ref{system throughput}), $\mathcal{R}_{\mathrm{tot}}$ is strictly concave with respect to $T$ in $(0,1)$, since it holds that
\begin{equation}
\begin{split}
&\frac{d^2\mathcal{R}_{\mathrm{tot}}}{dT^2}=-\frac{1}{\log(2)}\times\\
&\frac{(\eta \rho \sum_{n=1}^{N}g_n)^2}{T^3\ln(2)(1-\eta \rho \sum_{n=1}^{N}g_n+\frac{\eta \rho \sum_{n=1}^{N}g_n}{T})^2}<0.
\label{cancavity}
\end{split}
\end{equation}
Thus, the optimal value for $T$ in $(0,1)$ that maximizes $\mathcal{R}_{\mathrm{tot}}$ is unique and can be obtained through
\begin{equation}
\frac{d\mathcal{R}_{\mathrm{tot}}}{dT}=0.
\end{equation}
After some mathetmatical manipulations, the optimal value can be expressed as 
\begin{equation}
T^*=\frac{\eta \rho \sum_{n=1}^{N}g_n}{\eta \rho \sum_{n=1}^{N}g_n+\frac{\eta \rho \sum_{n=1}^{N}g_n-1}{W(\frac{\eta \rho \sum_{n=1}^{N}g_n-1}{e})}-1},
\label{optimal T}
\end{equation}
where $(\cdot)^*$ denotes a solution value and $W(x)$ returns the principal branch of the Lambert W function, also called omega function or product logarithm. This function is defined as the set of solutions of the equation $x=W(x)e^{W(x)}$ \cite{lambert}. Note that $W(x)$ can be easily evaluated since it is a built-in function in most of the well-known mathematical software packages as Matlab, Mathematica, etc. \cite{harvesting}. In the following, we describe two decoding order methods.

\subsection{Minimum Achievable Throughput Improvement with Descending Decoding Order}\label{maxthrdes}
Having optimized the achievable system throughput using (\ref{system throughput}), the next step is the selection of the decoding order of the users' messages. The simplest case is to adopt a fixed decoding order among users, that is, according to their indices. For fairness, the users' indices are assigned in a way that the values $g_n\forall N$  are sorted in descending order, i.e. $g_1\geq...\geq g_N$, since this allows decoding the weakest user's message without interference.  Therefore, this scheme increases both fairness and minimum achievable throughput, $\mathcal{R}_{\mathrm{min}}$ compared to other schemes with fixed decoding order, e.g. compared to ascending decoding order. 
 
\subsection{Minimum Achievable Throughput Optimization with Time-Sharing} 
Next, the time-sharing technique is utilized and optimized in order to improve the minimum achievable throughput among users, while the system throughput is maximized by setting $T=T^*$, where $T^*$ is given by (\ref{optimal T}). In contrast to fixed decoding order, the time-sharing technique has the benefit that, by proper selection of $\mathbold{\tau}$, any point of the capacity region can be achieved, and, thus, it can be exploited in order to improve fairness among the users. Also, as it has already been mentioned, the achievable system throughput is independent of the decoding order of the messages and, thus, the corresponding optimization scheme does not degrade the achievable system throughput. The resulting optimization problem can be written as 
\begin{equation}
\begin{array}{ll}
\underset{\mathbold{\tau}, \mathcal{R}_{\mathrm{min}}}{\text{\textbf{max}}}& \mathcal{R}_{\mathrm{min}} \\
\text{\textbf{s.t.}}&\mathrm{C}_n: \tilde{R}_n(T^*)\geq  \mathcal{R}_{\mathrm{min}},\,\forall{n\in\mathcal{N}},
\end{array}
\label{LP}
\end{equation}
where $\mathcal{N}=\{1,2,\ldots,N\}$ is the set of all users. 

The optimization problem in (\ref{LP}) is a \textit{linear programming} one and can be efficiently solved by well-known methods in the literature, such as simplex or interior-point method \cite{Boyd1}. In general, the worst-case complexity of linear problems is exponential in the dimensions of the problem, which for that in (\ref{LP}) is $(N+1)M$. 

A simple method to optimally apply the time-sharing technique, termed as full-space search, is to take into account all the permutations, i.e. to set, $M=N!$, in the user throughput definition in (\ref{with sharing}). Please, note that NOMA performs better for a small number of users, in which case the corresponding number of permutations might not be a barrier for the determination of the dynamic time-sharing configuration. For example, according to MUST scheme in LTE downlink only two users are grouped together for the implementation of NOMA \cite{3gpp}. Moreover, taking into account all possible permutations can be considered as a benchmark to other less complex schemes, which possibly exclude some permutations at the expense of a suboptimal configuration. 

Generalizing the above discussion, the full-space search is optimal, but inefficient when the number of users is large. To this end, a more efficient method will be discussed in the next subsection, while its effectiveness will be verified in the simulation results, where it will be compared with the full-space search. 

\subsection{A Greedy Algorithm for Efficient Time-Sharing}\label{greedy}
The complexity of the solution of the problem in (\ref{LP}) increases with the number of permutations, i.e. the inserted variables, which, in turn, increases considerably with the number of users. For a relatively small number of users, e.g. when $N=5$, $120$ permutations have to be taken into account. For the practical implementation of the time-sharing technique, considering such a number of permutations may be prohibitive. On the other hand, a priori exclusion of some permutations might cause severe degradation to the system performance in terms of minimum rate and fairness. For this purpose, in order to efficiently set the time-sharing configuration, an iterative method is proposed in this section. 

The main advantage of this method is that it finds and excludes some unnecessary permutations, without excluding the optimal configuration. In order to achieve this, instead of a priori considering all permutations, the set of permutations is dynamically constructed, while the corresponding time-sharing variables, $\tau_l$, are also optimized. 

The steps of the proposed greedy algorithm are discussed in detail below:
\begin{enumerate}
\item \textit{Initialization}: The users' indices are assigned in descending order with respect to $g_n$, such as in III.A, in order to construct the first permutation, i.e. $\mathbf{A}(1,j_{1,n})$. The achievable throughput of each user is calculated using (\ref{user rate}) and (\ref{N user rate}). 
\item \textit{Main loop (iteratively)}:
\begin{enumerate}[label=\roman*)]
\item The users' decoding order is determined according to the descending order of the throughput they achieve so far, for forming the new candidate permutation that will be inserted in $\mathbf{A}$. 
\item If the constructed permutation is not already included in $\mathbf{A}$, it is added in $\mathbf{A}$, while a new variable is inserted in $\mathbold{\tau}$. Adding new permutations with the described way gives the opportunity to the users that achieve small throughput to improve their rates, while achieving a balance between all users' rates, since the minimum achievable throughtput is never reduced. 
\item The linear optimization problem in (\ref{LP}) is solved for the updated $\mathbold{\tau}$.
\item  The new users' rates are calculated using (\ref{with sharing}).
\end{enumerate}
\item \textit{Convergence evaluation}: The main loop of the algorithm is repeated until the maximum number of iterations $K$ is reached, or a permutation is already included in $\mathbf{A}$. Please note that only new permutations are inserted in $\mathbf{A}$, because, otherwise, there would be two variables in $\mathbold{\tau}$ with exactly the same physical meaning.
\end{enumerate}

The above procedure can be summarized in Algorithm 1.

\begin{algorithm}\label{Alg}
\caption{: Greedy Algorithm for Efficient Time-Sharing Configuration}
\begin{algorithmic}[1]
\State \underline{\textbf{Initialization}} 
\State The users' decoding order variables $j_{1,n}$ are assigned in a way that the values 
$g_n\forall N$ are sorted in descending order. Thus, the first permutation is $\mathbf{A}(1,j_{1,n}),\,\forall n\in \mathcal{N} $, where 
$g_1\geq g_2\geq...\geq g_n\geq...\geq g_N$.
\State The users' rates are calculated using (\ref{user rate}) and (\ref{N user rate}).
\State Set $k=0$.
\State Set $\tilde{\mathcal{R}}_n[0]=\mathcal{R}_n,\, \forall n\in \mathcal{N}$.
\State \underline{\textbf{Main loop}} \Repeat 
\State Set $k=k+1$.
\State The users' decoding order variables $j_{k+1,n}$ are assigned in a way that the values of $\tilde{\mathcal{R}}_n[k-1]$ are in descending order. Thus, $\forall n,l \in \mathcal {N}$, \If {$\tilde{\mathcal{R}}_n[k-1]\leq \tilde{\mathcal{R}}_l[k-1]$}
\State Select $j_{k+1,n},j_{k+1,l}:j_{k+1,n}\geq j_{k+1,l}$. \EndIf
\State Update $\mathbf{A}$.
\If {$\exists n\in \mathcal {N}: \mathbf{A}(k+1,j_{k+1,n})\neq \mathbf{A}(m,j_{m,n}),\, \forall m\leq k$}
\State Solve (\ref{LP}), setting $M=k+1$.
\State Update the individual data rates $\tilde{\mathcal{R}}_n[k]$ using (\ref{with sharing}).
\EndIf
\Until $k=K$ or $\exists l\leq k: \mathbf{A}(k+1,j_{k+1,n})= \mathbf{A}(l,j_{l,n})\,\forall n \in \mathcal{N}$.
\end{algorithmic}
\end{algorithm}

\subsection{Examples}
In this subsection we present two examples for the cases of: i) similar channel conditions and, ii) significant difference between the channel values. In both examples, the number of users is fixed and equal to $N=2$, representing the simplest case, while the energy harvesting efficiency of each user is assumed to be $\eta_1=0.5$, and the amplifier's efficiency is $\eta_2=0.38$. All directional antenna gains are assumed to be equal to 0 dB (i.e. antenna gains are neglected). We assume a carrier center frequency of $470$ MHz, which will be used in the standard IEEE 802.11 for the next generation of Wi-Fi systems \cite{wifi, Kwan}. Furthermore, the TGn path loss model for indoor communication is adopted \cite{tgn,Kwan}, with the breakpoint distance being 5 m. More specifically, the path loss model that we use consists of the free space loss (slope of 2) up to the breakpoint distance and slope of 3.5 after the breakpoint distance. 

In both examples, for simplicity, we assume, $h_1=h_2=1$, $P_0=30$ dBm, and $N_0=-114$ dBm. For mathematical clarity, $d_n$ denotes the distance between the BS and the $n$-th user.

\begin{figure}
\centering
\includegraphics[width=\columnwidth]{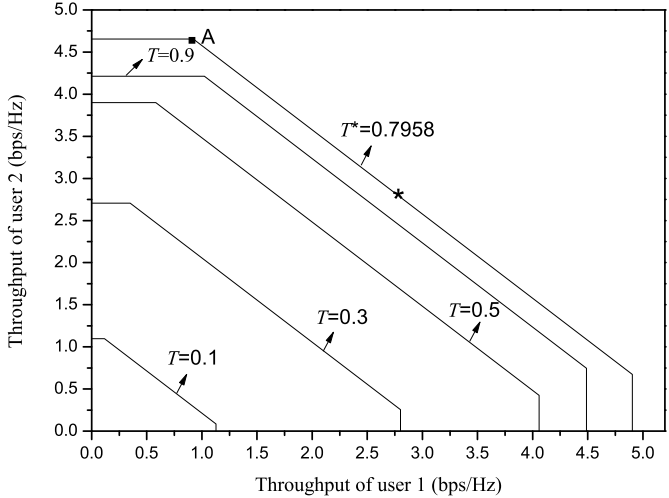}
\caption{Example 1: Achievable throughput region.}
\label{Fig2}
\end{figure}

\textit{Example 1 (Similar distance)}:
For the first example, it is assumed that $d_1=9.9$ m and $d_2=10.1$ m, which corresponds to ${L}_{01}=2.4067 \cdot 10^{-6}$ and ${L}_{02}=2.156 \cdot 10^{-6}$ . This example is representative of the case of two users located in a similar distance from the BS. Fig. \ref{Fig2} depicts the capacity regions for the two users for different choices of $T$, as well as, for the optimal value of $T$, which is $T^*=0.7958$. Fixed descending decoding order with respect to the channel values results in different achievable throughput of the two users, degrading the throughput of the user with the best channel conditions, i.e. $\mathcal{R}_1=0.6727$ bps/Hz and $\mathcal{R}_2=4.90535$ bps/Hz. This point corresponds to the upper-left corner (A) of the capacity region.

Interestingly, the capacity region which is formed when $T=0.7958$, includes a set of solutions that dominates, in terms of both achievable system throughput and minimum user throughput, any other set of solutions, imposed by the capacity region formed by any other value of $T$. This is an important observation, taking into account that any point of the capacity region can be achieved with proper time-sharing configuration. In this example, by choosing $\tau_1=0.4688$ and $\tau_2=0.5312$, the users' achievable throughput becomes $\tilde{\mathcal{R}}_1=\tilde{\mathcal{R}}_2=2.7891$ bps/Hz. This configuration corresponds to the point that is marked with asterisk.

\begin{figure}
\centering
\includegraphics[width=\columnwidth]{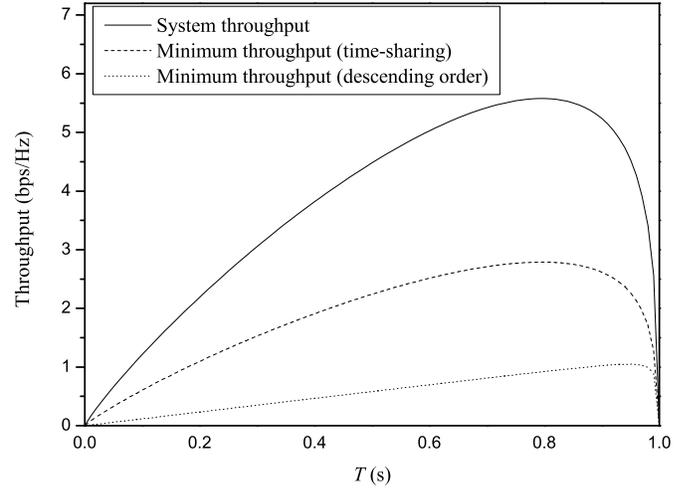}
\caption{Example 1: System throughput and minimum-throughput.}
\label{Fig3}
\end{figure}

In Fig. \ref{Fig3}, the minimum throughput between the two users, with respect to the value of T is depicted, while the achievable system throughput of the users is also depicted as a reference. It is shown that by using and optimizing the time-sharing, the value of $T$ that maximizes the system throughput also maximizes the minimum user throughput. This is because $\mathcal{R}^*_\mathrm{min}=\frac{\mathcal{R}^*_{\mathrm{tot}}}{2}$. On the other hand, it is illustrated that when the fixed descending decoding order is chosen, then the value of $T$ that maximizes the minimum user throughput is higher than $T^*$, which corresponds to a lower value of system throughput. However, as it will be shown in the next example, the solution that maximizes the system throughput does not always maximize the minimum-throughput, even with proper time-sharing configuration.

\begin{figure}
\centering
\includegraphics[width=\columnwidth]{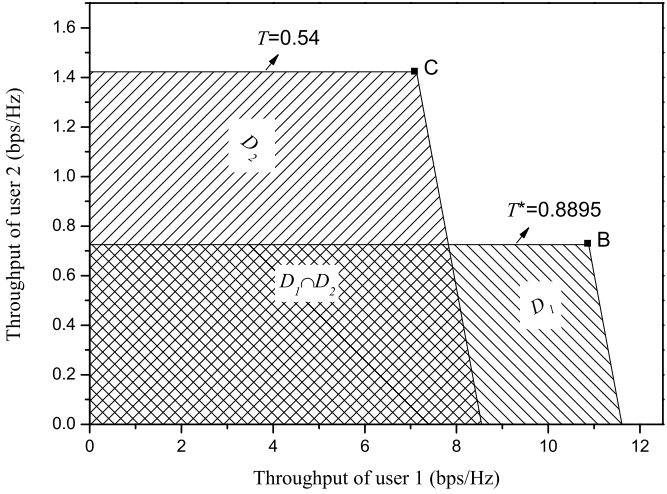}
\caption{Example 2: Achievable throughput region.}
\label{Fig4}
\end{figure}

\textit{Example 2 (The ``double near-far" problem)}:
For this example, the distances between the users and the BS are chosen in a way that $\mathcal{L}_{01}\gg \mathcal{L}_{02}$, i.e. $d_1=6$ m and $d_2=14$ m, such as that ${L}_{01}=3.7808 \cdot 10^{-5}$ and ${L}_{02}=2.5786 \cdot 10^{-7}$. This configuration is a representative of the ``double near-far" phenomenon, which appears when a user far from the BS receives a smaller amount of wireless energy than a nearer user, while it needs to transmit with more power \cite{Ju}. When NOMA is used, this phenomenon directly affects the capacity region, as it is evident from Fig.\ref{Fig4}. As it can be observed, the value of $T$ that maximizes the system throughput is equal to $T^*=0.8895$. 

When descending decoding order (with respect to the channel values) is utilized, then the achievable throughput values are $\mathcal{R}_1=10.8823$ bps/Hz and $\mathcal{R}_2=0.7251$ bps/Hz. This point corresponds to the corner (B) of the region $D_1$. It is remarkable that the set of solutions included in the capacity region that is formed when $T=T^*$ does not dominate any other set of solutions both in terms of system throughput and minimum user throughput, e.g. $D_2$ is not a subset of $D_1$. This means that the time-sharing technique cannot improve the minimum throughput, which, however can be improved by a different selection of $T$. For example, when $T=0.54$ the users' throughput values become $\mathcal{R}_1=7.1242$ bps/Hz and $\mathcal{R}_2=1.4223$ bps/Hz. This point corresponds to the corner (C) of the region $D_2$. However, this selection does not maximize the system throughput.

The above examples have illustrated the trade-off between the minimum user throughput and system throughput. Minimum throughput maximization is an important problem, especially for communication networks where the users have similar quality of service requirements. For this reason, equal individual data rate maximization will be further discussed and optimized in the next sections.

\section{equal Individual Data Rate Maximization with Descending Decoding Order}
In this section, we aim to maximize the equal individual data rate, i.e.  the minimum user throughput in the case where all users aim to transmit with an equal rate, $\mathcal{R}_\mathrm{eq}$, and, thus, $T$ can be adjusted accordingly. Fixed descending decoding order is assumed for simplicity. Next, we present and efficiently solve the corresponding optimization problem.
\subsection{Problem Formulation}
The problem of equal individual data rate maximization, when the message of the users with the best channel conditions is decoded first, can be written as:
\begin{equation}
\begin{array}{ll}
\underset{T}{\text{\textbf{max}}}& \mathcal{R}_\mathrm{eq} \\
\text{\textbf{s.t.}}&\textit{C}_n: R_n\geq  \mathcal{R}_\mathrm{eq},\,\forall{n\in\mathcal{N}}, \\
&\mathrm{C}_{N+1}:0<T<1.
\end{array}
\label{opt_dec}
\end{equation}
The first $N$ constraints of the optimization problem in (\ref{opt_dec}) are strictly concave since
\begin{equation}
\begin{split}
&\frac{d^2R_{n}}{dT^2}=-\frac{1}{\log(2)}\times\\
&\frac{a_n\left(2b_n\left(b_n\left(1-T\right)+T\right)+a_n\left(2b_n\left(1-T\right)+T\right)\right)}{\left(a_n\left(-1+T\right)+b_n\left(-1+T\right)-T\right)^2\left(b_n+T-b_nT\right)^2}<0,
\end{split}
\label{concavity}
\end{equation}
where $\mathbf{A}_n=\eta\rho g_n$, $b_n=\eta\rho\sum_{j=n+1}^Ng_j$, and $b_N=0$. Also, the objective function, as well as the $(N+1)$-th constraint, are linear, and therefore  (\ref{opt_dec}) is a convex optimization problem, which can be solved by standard numerical methods such as a combination of interior point methods and bisection method.
However, we use dual-decomposition, which, apart from giving physical insights into the structure of the problem, proves to be extremely efficient, since, given the Lagrange multipliers (LMs), the optimal $T$ can be directly calculated.  More importantly, using the adopted method, it is guaranteed that the optimal solution can be obtained in polynomial time \cite{Boyd1}. Also, note that by using dual-decomposition our work is directly comparable to \cite{Ju}, in terms of complexity of the provided solution for the equal individual data rate maximization problem, among others.

\subsection{Dual Problem Formulation and Solution of (\ref{opt_dec})}
In this subsection, the optimization problem (\ref{opt2}) is solved by \textit{Lagrange dual decomposition}. In order to handle the linear objective function in (\ref{opt_dec}), we replace it with $\ln(R_\mathrm{eq})$, without affecting the convexity. Since the primal problem is convex and satisfies the Slater's condition qualifications, strong duality holds, i.e., solving the dual is equivalent to solving the primal problem \cite{Boyd1}. In order to formulate the dual problem, the Lagrangian of (\ref{opt_dec}) is needed, which is given by

\begin{equation}
\begin{split}
&\mathcal{L}(\mathbold{\lambda},T,\mathcal{R}_\mathrm{eq})=\ln(\mathcal{R}_\mathrm{eq})\\
&+\sum_{n=1}^N\lambda_n\left(T\log_2\left(1+\frac{a_n}{b_n+\frac{T}{1-T}}\right)-\mathcal{R}_\mathrm{eq}\right),\\
\end{split}
\label{lagrangian1}
\end{equation}
where $\lambda_n\geq 0$ is the Lagrange multiplier (LM), which corresponds to the constraint $\mathrm{C}_n$ and $\mathbold{\lambda}$ is the Lagrange multiplier vector with elements $\lambda_n$. The constraint $\mathrm{C}_{N+1}$ is absorbed into the Karush-Kuhn-Tucker (KKT) conditions, and is presented in detail in the next subsection.

The dual problem is now given by \begin{equation}
\underset{\mathbold{\lambda}}{\text{\text{min}}}\,\,\,\,\underset{T,\mathcal{R}_\mathrm{eq}}{\text{\text{max}}}\,\,\mathcal{L}(\mathbold{\lambda},T,\mathcal{R}_\mathrm{eq}).
\label{dual problem 1}
\end{equation}

According to the KKT conditions, the optimal values of $\mathcal{R}_{\mathrm{eq}}$ and $T$ are given by 
\begin{equation}
\mathcal{R}_{\mathrm{eq}}^*=\frac{1}{\sum_{n=1}^N\lambda_n}
\label{optimal_rate}
\end{equation}
and
\begin{equation}
\begin{split}
&T^*=\left[\frac{x^*}{1+x^*}\right]_{\epsilon}^{\varepsilon},
\end{split}
\label{optimalT2}
\end{equation}
where $[\cdot]_x^y=\text{min}(\text{max}(\cdot,x),y)$, $\epsilon\rightarrow 0^+$, $\varepsilon\rightarrow 1^-$, and $x^*$ is the solution of the following equation:
\begin{equation}
\sum_{n=1}^N\lambda_n\log(1+\frac{a_n}{b_n+x})=\sum_{n=1}^N\lambda_n\frac{a_n(x+x^2)}{(b_n+x)^2+a_n(b_n+x)}.
\end{equation}

The dual problem in (\ref{dual problem 1}) can be solved iteratively. In each iteration, the optimal $\mathcal{R}_{\mathrm{eq}}$ and $T$ are calculated for a fixed LM vector, $\mathbold{\lambda}$, using (\ref{optimal_rate}) and (\ref{optimalT2}), while $\mathbold{\lambda}$ is then updated using the gradient method as follows
\begin{equation}
\begin{split}
&\lambda_n[t+1]=\left[\lambda_n[t]-\hat{\lambda}_n[t]\times\right.\\
&\left.\left(T\log_2\left(1+\frac{a_n}{b_n+\frac{T}{1-T}}\right)-\mathcal{R}_\mathrm{eq}\right)\right]^+,\,\forall n \in \mathcal{N},
\end{split}
\end{equation}
where $t$ is the iteration index, $\hat{\lambda}_n,\,n\in\mathcal{N}$ are positive step sizes, and $[\cdot]^+=\min(\cdot,0)$. Since (\ref{opt_dec}) is concave, it is guaranteed that the iterations between the two layers converge to the optimal solution if the size of the chosen step satisfies the infinite travel condition \cite{Boyd2}
\begin{equation}
\sum_{t=1}^{\infty}\hat\lambda_n[t]=\infty,\,\forall n\in\mathcal{N}.
\end{equation}

As it can be observed from the solution in (\ref{optimal_rate}), the equal individual data rate is inversely proportional to the sum of the LMs. This result is consistent with the physical interpretation of the  LMs, which are indicative of how active the corresponding constraints are, depicting the impact of the weakest users, via the violated constraints, on the optimal value. If $\lambda^*_n$ is  small it means that the effect of the $n$-th constraint on the determination of $\mathcal{R}_{\mathrm{eq}}^*$ in (\ref{optimal_rate}), as well as of $x^*$ in (\ref{optimalT2}), is not significant. On the other hand, if $\lambda_n^*$ is large it means that if the constraint is loosened or tightened a bit, the effect on $\mathcal{R}_{\mathrm{eq}}^*$ will be great \cite{Boyd1}. In this case, the throughput of the $n$-th user is in high priority when optimizing the time that is dedicated to energy transfer.

\section{Equal Individual Data Rate Maximization with Time-Sharing}
In contrast to the previous section, where we assumed fixed descending decoding order, we aim to maximize the equal individual data rate, while utilizing the time-sharing technique. For this purpose, $T$ as well as the time-sharing configuration need to be optimized. Please note that in contrast to the time-sharing configuration discussed in section \ref{maxthrdes}, the solution provided in this section does not necessarily maximize the system throughput. 
 
\subsection{Problem Formulation and Solution}
Next, the indices of the users are ordered according to $g_1\geq g_2\geq\ldots\geq g_N$, however, the order of decoding depends on the time-sharing. 
Taking into account the above considerations, the problem of equal individual data rate maximization can be formulated as
\begin{equation}
\begin{array}{ll}
\underset{\mathbold{\tau},T,\mathcal{R}_\mathrm{eq}}{\text{\textbf{max}}}& \mathcal{R}_\mathrm{eq} \\
\text{\textbf{s.t.}}&\textit{C}_n: \tilde{R}_n\geq  \mathcal{R}_\mathrm{eq},\,\forall{n\in\mathcal{N}}, \\
&\mathrm{C}_{N+1}:0<T<1.
\end{array}
\label{opt}
\end{equation}

Obviously, the optimization problem in (\ref{opt}) is a non-convex one, due to the coupling of the variables $T$ and $\tau$. We note that there is no standard approach for solving non-convex optimization problems in general. In order to derive an efficient and optimal time allocation method for the considered problem we take into account the following observations.

\begin{remark}\label{problem seperation}
A selection for $T$ corresponds to a specific capacity region for the set of users $\mathcal{N}$, where the time-sharing technique can also be used. On the other hand, each of the points of this capacity region corresponds to a different selection of $\tau$. As we have already mentioned, with proper selection of the time-sharing variables, any point of the capacity region can be achieved.
\end{remark}

Taking into account Remark \ref{problem seperation}, for a given time $T$, the achievable rate region is defined by the inequalities
\begin{equation}
\begin{split}
\tilde{R}_n(T)&\leq T\log_2\left(1+\tfrac{\eta\rho g_n\left(1-T\right)}{T}\right),\,\forall n\in\mathcal{N}\\
\sum_{n\in\mathcal{M}_k} \tilde{R}_n(T)&\leq T\log_2\left(1+\tfrac{\eta\rho \left(1-T\right)\sum g_n}{T}\right),\,\forall k: \mathcal{M}_k\subseteq\mathcal{N},
\end{split}
\end{equation}
where the second inequality holds for any sum set, $\mathcal{M}_k\subseteq\mathcal{N}$. Now, suppose that the BS cancels all other users' messages, except the user with the weakest link. In this case it is desired that its throughput is at least equal to the final achievable $\mathcal{R}_\mathrm{eq}$, i.e.
\begin{equation}
\underbrace{T\log_2\left(1+\frac{\eta\rho\left(1-T\right)g_N}{T}\right)}_{R_N}\geq\mathcal{R}_\mathrm{eq}.
\end{equation}
Accordingly, for the two weakest users, that is for $n=N$ and $n=N-1$, their sum-throughput is maximized when the BS cancels out all other users' messages, while one of the two messages is also canceled. Since they can allow time-sharing for the time that each user's message will be canceled, for the sum of the throughput of these two users it must hold that
\begin{equation}
\underbrace{T\log_2\left(1+\frac{\eta\rho\left(1-T\right)\sum_{n=N-1}^Ng_n}{T}\right)}_{R_{N-1}+R_N}\geq2\mathcal{R}_\mathrm{eq}.
\end{equation}

Following the same strategy for all other users, it yields that $\mathcal{R}_{\mathrm{eq}}$ is bounded by the following set of inequalities
\begin{equation}
\mathcal{R}_{\mathrm{eq}} \leq \frac{T\log_2\left(1+\frac{\eta \rho\sum_{i=n}^{N}g_i}{\frac{T}{1-T}}\right)}{(N+1-n)},\,\forall{n\in\mathcal{N}},
\end{equation}
in which $\mathbold{\tau}$ does not appear.
Consequently, the optimization in \eqref{opt} can be optimally solved by reducing it into two disjoint problems, after minimizing the initial
search space. These optimization problems are:

\emph{Problem 1: Optimization of $T$}
\begin{equation}
\begin{array}{ll}
\underset{T}{\text{\textbf{max}}}& \mathcal{R}_\mathrm{eq} \\
\text{\textbf{s.t.}}&\textit{C}_n: T\log_2\left(1+\frac{\eta \rho\sum_{i=n}^{N}g_i}{\frac{T}{1-T}}\right)\geq\\
&\quad\quad\quad   (N+1-n)\mathcal{R}_\mathrm{eq},\,\forall{n\in\mathcal{N}}, \\
&\mathrm{C}_{N+1}:0<T<1,
\end{array}
\label{opt2}
\end{equation}

\emph{Problem 2: Calculation of the time-sharing vector $\mathbold{\tau}$}
\begin{equation}
\begin{array}{ll}
{\text{\textbf{find}}}& \mathbold{\tau} \\
\text{\textbf{s.t.}}&\mathrm{C}_n: \tilde{R}_n(T^*)\geq  \mathcal{R}^*_\mathrm{eq},\,\forall{n\in\mathcal{N}}.
\end{array}
\label{LP2}
\end{equation}

In the above, $\mathcal{R}^*_\mathrm{eq}$, denotes the optimal solution for $\mathcal{R}_\mathrm{eq}$, which is calculated by solving Problem 1. The solution of Problem 2 is calculated after the solution of Problem 1. Please note that, when solving Problem 2, since $T^*$ and $\mathcal{R}^*_\mathrm{eq}$ have already been fixed, this is a linear optimization problem, with similar structure to (\ref{LP}). Thus, it can be solved by utilizing the same linear programming methods or  by using Algorithm 1. On the other hand, Problem 1 is jointly concave with respect to $T$ and $\mathcal{R}_\mathrm{eq}$, and satisfies Slater's constraint qualification. Thus, it is a convex optimization problem, which can be solved by following similar steps as in the solution of (\ref{opt_dec}).

\subsection{Solution of Problem 1}
In this subsection, the optimization problem (\ref{opt2}), i.e. Problem 1, is solved by \textit{Lagrange dual decomposition}. The Lagrangian of Problem 1, after replacing the initial objective function with $\ln(\mathcal{R}_{\mathrm{eq}})$, is given by
\begin{equation}
\begin{split}
\mathcal{L}(\mathbold{\lambda},T,\mathcal{R}_\mathrm{eq})&=\ln(\mathcal{R}_\mathrm{eq})+\sum_{n=1}^N\mu_n\times\\
&\left(\frac{T\log_2\left(1+\frac{c_n}{\frac{T}{1-T}}\right)}{d_n}-\mathcal{R}_\mathrm{eq}\right),
\end{split}
\label{lagrangian}
\end{equation}
where $c_n=\eta \rho\sum_{i=n}^{N}g_i$, $d_n=N+1-n$, $\mu_n\geq 0$ is the Lagrange multiplier, which corresponds to the constraint $\mathrm{C}_n$ and $\mathbold{\mu}$ is the Lagrange multiplier vector with elements $\mu_n$. 

The dual problem is now given by 
\begin{equation}
\underset{\mathbold{\mu}}{\text{\text{min}}}\,\,\,\,\underset{T,\mathcal{R}_\mathrm{eq}}{\text{\text{max}}}\,\,\mathcal{L}(\mathbold{\mu},T,\mathcal{R}_\mathrm{eq}).
\label{dual problem}
\end{equation}

%
%

The dual problem in (\ref{dual problem}) can be iteratively solved, as we did in problem (\ref{dual problem 1}). In each iteration, the optimal values of $\mathcal{R}_{\mathrm{eq}}$ and $T$ are given by 
\begin{equation}
\mathcal{R}_{\mathrm{eq}}^*=\frac{1}{\sum_{n=1}^N\mu_n}
\end{equation}
and
\begin{equation}
\begin{split}
T^*=\left[T\in\mathbb{R}:\sum_{n=1}^{N}\frac{\mu_n}{d_n}\left(\ln\left(1+\frac{c_n}{T}-
c_n\right)\right.\right.\\
\left.\left.-\frac{c_n}{T(1-c_n)+c_n}\right)=0\right]_{\epsilon}^{\varepsilon}.
\end{split}
\label{optimalT3}
\end{equation}

Furthermore, the LMs can be updated as follows
\begin{equation}
\begin{split}
\mu_n[t+1]=&\left[\mu_n[t]-\hat{\mu}_n[t]\left(\frac{T\log_2\left(1+\frac{c_n}{\frac{T}{1-T}}\right)}{d_n}-\mathcal{R}_{\mathrm{eq}}\right)\right]^+,\\
&\,\forall n\in\mathcal{N},
\end{split}
\end{equation}
where $\hat{\mu}_n,\,n\in\mathcal{N}$ are positive step sizes.

\section{Simulation Results and Discussions}

For the simulations, we assume that the users are uniformly distributed in a ring-shaped surface, with $r_{c1}=5$ m and $r_{c2}=20$ m being the radii of the inner and the outer circle, respectively, while the BS is located at the center of the circles.  The path loss model, as well as the energy harvesting and the amplifier's efficiency are set according to section \ref{greedy}. All statistical results are averaged over $10^5$ random channel realizations.  The receiver of the BS is assumed to have a white power spectral density of $-174$ dBm/Hz, while all directional antenna gains are assumed to be 7.5 dB, and the available bandwidth is considered to be 1 MHz. Finally, all permutations are taken into account when optimizing the time-sharing configuration, i.e. $M=N!$, unless stated otherwise.

The main focus of the simulation results is the comparison of the performance among the proposed optimization schemes, i.e. (a)-(d). 
Next, the resulting solutions by the aforementioned optimization schemes are compared in terms of system or user throughput, portion of time that is dedicated to energy harvesting, energy efficiency and user fairness. Also, they are presented against the corresponding results of the baseline orthogonal (TDMA) scheme, which is considered in \cite{Ju}. Next, for the readers' convenience, we use the following notations regarding the comparison with the TDMA approach \cite{Ju}:
\begin{enumerate} [label=\Alph*.]
\item System throughput maximization. 
\item Equal individual data rate maximization.
\end{enumerate}
Note that, in \cite{Ju}, case B is referred to as ``common-throughput''.

Moreover, the convergence speed of the greedy algorithm, i.e. Algorithm 1 is also evaluated.

\subsection{Throughput Comparison}
\begin{figure}
\centering
\includegraphics[width=\columnwidth]{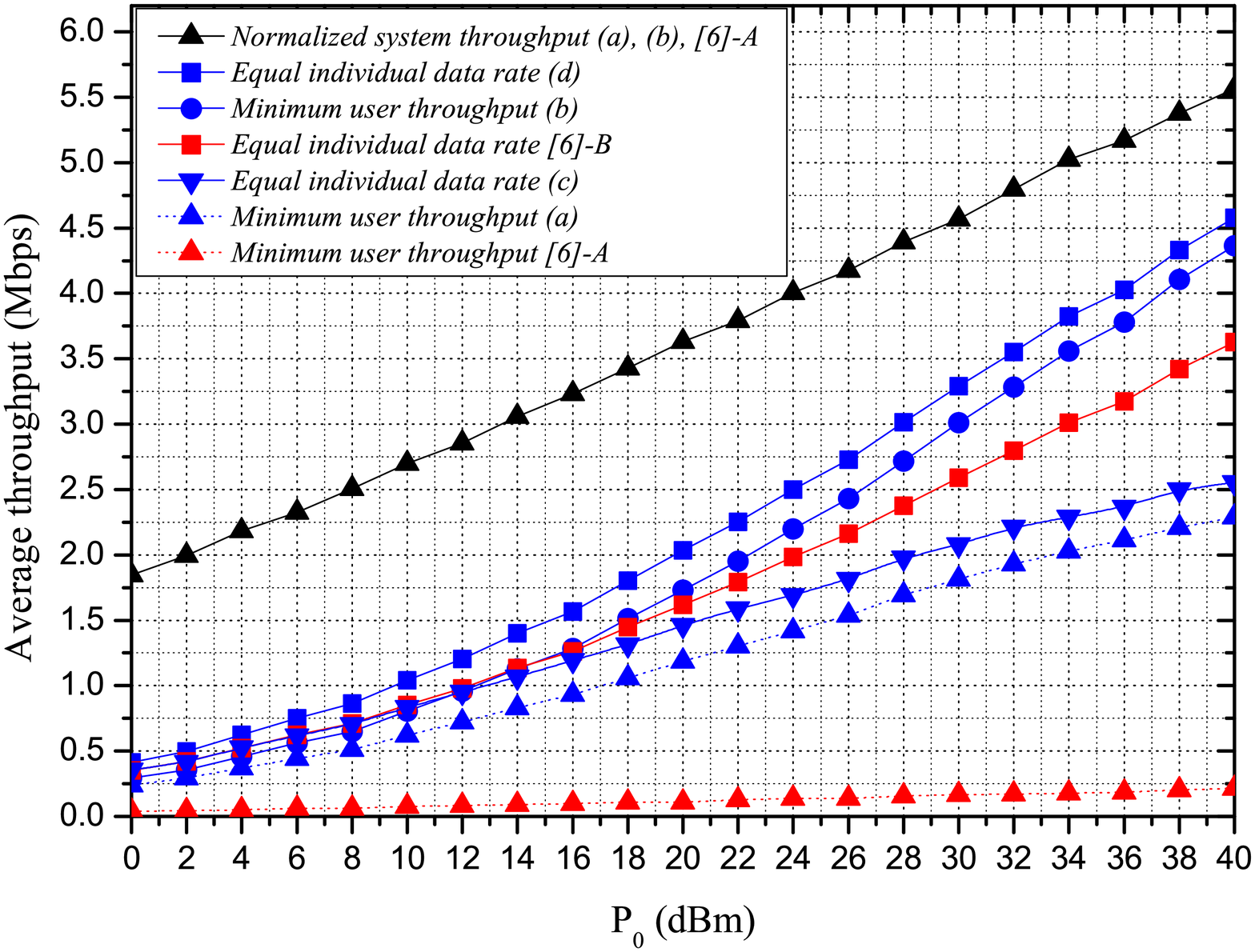}
\caption{Comparison of average throughput, when $N=3$.}
\label{Fig5}
\end{figure}

In Fig. \ref{Fig5}, the average throughput of the weakest user that is achieved by all methods discussed in this paper, is illustrated and compared for the case of $N=3$. More specifically, Fig. \ref{Fig5} includes: i) the minimum user throughput that NOMA with fixed decoding order and TDMA can achieve, when maximizing the system throughput, ii) the equal individual data rate that NOMA and TDMA can achieve, and iii) the minimum user throughput that NOMA achieves without reducing the system throughput, employing time-sharing. For reference, the normalized system throughput that is achieved in this case (i.e., the system throughput divided by the number of users, $\frac{\mathcal{R}_{\mathrm{tot}}}{N}$), is also depicted. It is evident that both NOMA and TDMA achieve the same normalized system throughput, however in this case, the application of the proposed NOMA scheme results in a notable increase of the minimum user throughput, for the whole range of $P_0$, even when time-sharing is not used. 

As it can be observed, NOMA performs better when combined with time-sharing. This is because when fixed descending decoding order is used, only the corner points of the capacity region can be achieved. Therefore, the rates of the users with weaker channel conditions are improved at the expense of the rates of the rest of the users, while the minimum user throughput is not necessarily the one of the user with the weakest channel conditions. However, this is not the case when time-sharing is used and optimized, increasing the degrees of freedom, since any point of the capacity region can be achieved. Consequently, when time-sharing is applied, the minimum throughput that NOMA achieves is larger, even compared to the equal individual data rate achieved by TDMA, for the medium and high region of $P_0$. Moreover, when maximizing the equal individual data rate, NOMA with time-sharing clearly outperforms TDMA, and the equal rate is higher for the whole range of the transmit power values of the BS, and especially in the high $P_0$ region. 

\begin{figure}
\centering
\includegraphics[width=\columnwidth]{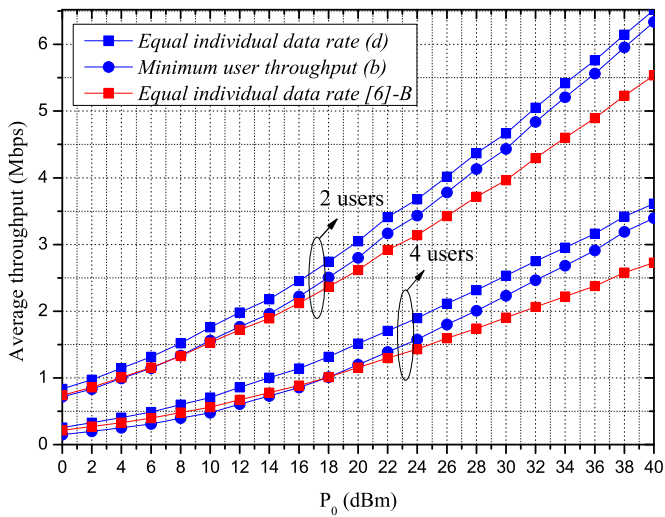}
\caption{Comparison of (b), (d) and [6]-B in terms of average throughput of the weakest user.}
\label{Fig6}
\end{figure}

Fig. \ref{Fig6} depicts the impact of the number of users on the system's performance. It can be easily observed that, as the number of users increases, both the equal individual data rate and the minimum achievable throughput that NOMA achieves decrease. However, the first is always higher than that achieved by TDMA. Furthermore, as the number of users increases, the difference between equal individual data rate and minimum user throughput that NOMA achieves, when maximizing the equal individual data rate and the system throughput, respectively, also increases. Thus, when $N=2$, maximizing the system throughput has a less significant impact on the minimum individual data rate, compared to the case when $N=4$.

\subsection{Trade-off Between Energy Harvesting and Information Transmission}
\begin{figure}[t!]
\centering
\includegraphics[width=\columnwidth]{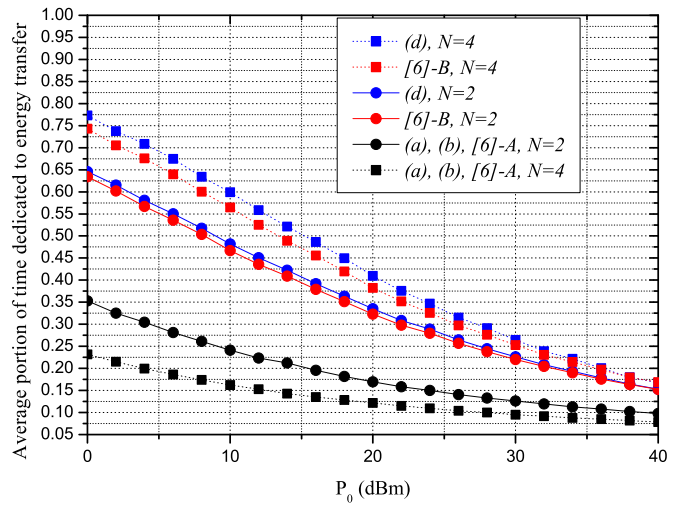}
\caption{Comparison among (a), (b), (d), [6]-A, and [6]-B in terms of portion of time dedicated to energy transfer.}
\label{Fig7}
\end{figure}

In Fig. \ref{Fig7}, the charging time is depicted when the system throughput and the equal individual data rate are maximized, for $N=2$ and $N=4$. As it can be observed, when the aim is to maximize the system throughput and the number of users increases, the portion of time dedicated to energy transfer is reduced. This happens because the system throughput is mainly affected by the individual data rates of the users with good channel conditions, the average number of which increases with $N$, since uniform distribution has been assumed for the users' locations. Moreover, the users with good channel conditions tend to prefer higher values of $T^*$, compared to those with worse channel conditions, in order to improve their individual data rates. In other words, they have enough energy to transmit and, as a result, their sensitivity to the resource of time dedicated to information transmission increases. 

On the other hand, when the equal individual data rate is maximized, the weakest user, i.e. the one with the worst channel conditions, must have enough energy supply to achieve the equal individual data rate. In this case, as the number of users increases, the portion of time dedicated to energy transfer also increases. Moreover, NOMA dedicates slightly more time to energy harvesting compared to TDMA. The reason for this is that NOMA exploits more efficiently the time dedicated to information transmission, requiring less time for achieving the same equal data rate with TDMA. However, when the time dedicated to energy transfer increases, the energy consumption to the BS's side also increases. The last observation motivates the comparison of the two schemes, i.e. NOMA and TDMA, in terms of energy efficiency.

\subsection{Energy Efficiency}
\begin{figure}
\centering
\includegraphics[width=\columnwidth]{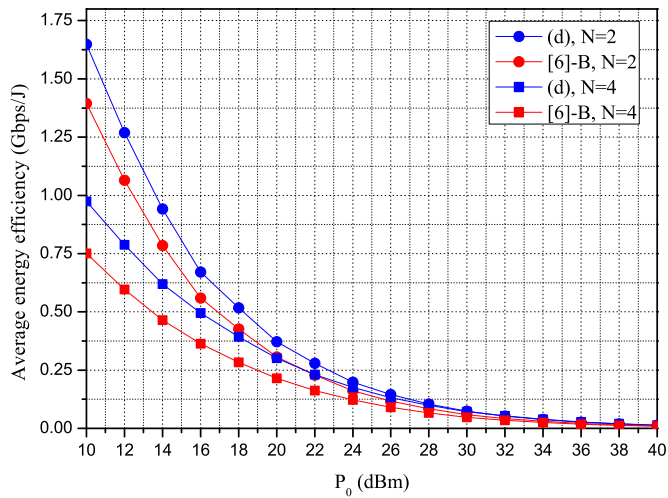}
\caption{Comparison of average energy efficiency between (d) and [6]-B.}
\label{Fig8}
\end{figure}

The efficiency of the energy transmitted by the BS, denoted by $\mathcal{E}$, when equal transmission rate is required among users, is defined as
\begin{equation}
\mathcal{E}=\frac{N\mathcal{R}_{\mathrm{eq}}}{P_0(1-T)}.
\end{equation}
In Fig. \ref{Fig8}, NOMA and TDMA are compared in terms of energy efficiency. It is remarkable that although NOMA dedicates more time to energy harvesting when compared to TDMA, i.e. more energy is transmitted, it achieves higher energy efficiency for the whole range of $P_0$. This is because NOMA achieves much higher individual rates compared to TDMA. Another important observation is that the energy efficiency is decreased considerably, when the value of $P_0$ is increased. Consequently, there is a clear trade-off between equal individual data rate and energy efficiency.

\subsection{Fairness Comparison}
\begin{figure}
\centering
\includegraphics[width=\columnwidth]{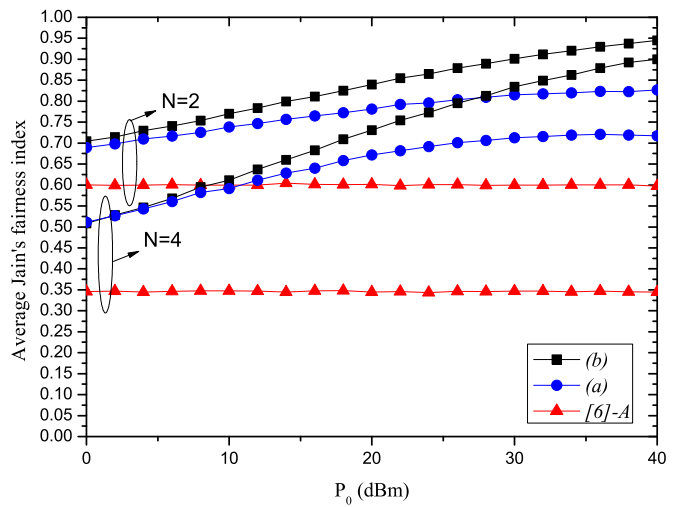}
\caption{Jain's fairness index comparison.}
\label{Fig9}
\end{figure}

In order to fairly compare the two schemes (NOMA and TDMA) in Fig. \ref{Fig9}, we use the Jain's fairness index, $\mathcal{J}$,  which is given by \cite{Tafazoli}
\begin{equation}
\mathcal{J}=\frac{(\sum_{n=1}^NR_n)^2}{N\sum_{n=1}^NR_n^2}.
\end{equation}
Note that Jain's fairness index is bounded between 0 and 1, with unitary value indicating equal users' rates. It is seen in Fig. \ref{Fig9} that NOMA provides more fairness compared to TDMA, for the whole range of $P_0$. Also note that the three illustrated schemes achieve the same system throughput, for the same number of users.

\subsection{Convergence of the Greedy Algorithm}
\begin{figure}
\centering
\includegraphics[width=\columnwidth]{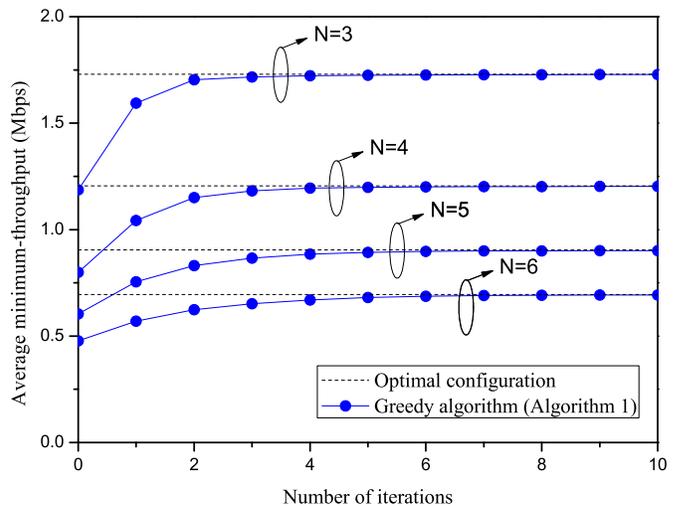}
\caption{Evaluation of the convergence speed of the greedy algorithm.}
\label{Fig10}
\end{figure}

Fig. \ref{Fig10} illustrates the evolution of the average minimum user throughput when the proposed greedy algorithm is used for the time-sharing configuration. In particular, we focus on the convergence speed of the proposed algorithm for $P_0=20$ dBm and $N=3,4,5,6$. The dashed lines denote the minimum user throughput for each case study. It is observed that the proposed iterative algorithm converges to the optimal value within $N+1$ iterations. Thus, the proposed technique reduces the maximum number of permutations that have to be considered by the full-space search from $N!$ (when all permutations are considered) to $N+2$, rendering the time-sharing technique suitable for practical implementation.

\vspace{0.05 in}
\section{Conclusions}
In this paper, we have studied time-allocation methods in order to maximize the individual data rates and to improve fairness in wireless powered communication systems with NOMA. All formulated optimization problems were solved by using linear programming methods and convex optimization tools. Also we have compared the proposed scheme with the case that the energy harvesting nodes utilize TDMA, which was considered as a baseline. Extensive simulation results have shown that the proposed scheme outperforms the baseline, in terms of throughput and fairness. Finally, they reveal an interesting dependence among system throughput, minimum individual data rate, and harvested energy.

\balance

\end{document}